\documentclass[a4paper]{jpconf}
\usepackage{graphicx}
\begin{document}

\title{Microscopic DC-TDHF study of heavy-ion potentials 
and fusion cross sections}

\author{V E Oberacker$^1$, A S Umar$^1$ and R Keser$^{2}$}

\address{$^1$ Department of Physics and Astronomy, Vanderbilt University, Nashville, TN 37235, USA}
\address{$^2$ RTE University, Science and Arts Faculty, Department of Physics, 53100, Rize, TURKEY}

\ead{volker.e.oberacker@vanderbilt.edu}


\begin{abstract}
We study heavy-ion fusion reactions at energies near the Coulomb barrier, in particular
with neutron-rich radioactive ion beams. Dynamic microscopic calculations
are carried out on a three-dimensional lattice using the Density-Constrained 
Time-Dependent Hartree-Fock (DC-TDHF) method. New results are presented for the
$^{132}$Sn+$^{40}$Ca system which are compared to $^{132}$Sn+$^{48}$Ca studied
earlier. Our theoretical fusion cross-sections agree surprisingly well with recent
data measured at HRIBF. We also study the near- and sub-barrier fusion of $^{24,16}$O
on $^{12}$C which is important to determine the composition and heating of the crust
of accreting neutron stars.
\end{abstract}


\section{Introduction}

The calculation of heavy-ion interaction potentials is of fundamental
importance for the study of fusion reactions between stable and
neutron-rich nuclei, and for the study of superheavy element production.
We have developed a fully microscopic method to extract ion-ion potentials
directly from the Time-Dependent Hartree-Fock (TDHF) time-evolution of the
nuclear system. The only input is the effective NN interaction (Skyrme), and there are
no adjustable parameters. 

Radioactive ion beam facilities have opened up the possibility to study fusion
reactions of neutron-rich nuclei. In many cases, the dynamics
of the neutron rich skin of these nuclei enhances the sub-barrier fusion cross
section over that predicted by a simple static barrier penetration model, but
in some cases suppression of fusion is also observed. 
Very recently, at the HRIBF facility a series of experiments has been carried out
with radioactive $^{132}$Sn beams on $^{40,48}$Ca targets~\cite{KR12}. 
Also, at the GANIL-SPIRAL facility a reaccelerated beam of $^{20}$O was used to
measure near- and sub-barrier fusion with a  $^{12}$C target~\cite{desouza}. 

The time-dependent Hartree-Fock (TDHF) theory provides a useful foundation for a
fully microscopic many-body theory of large amplitude collective
motion~\cite{Ne82,Cus85a} including deep-inelastic and fusion reactions.
Recently it has become feasible, for the first time, to perform TDHF calculations on a
3D Cartesian grid without any symmetry restrictions
and with much more accurate numerical methods~\cite{UO06,KS10,Si11,GM08,DD-TDHF}.
At the same time the quality of effective interactions has also been substantially
improved~\cite{CB98,Klu09a,KL10}.
During the past several years, we have developed the DC-TDHF method for calculating
heavy-ion potentials~\cite{UO06a}, and we have applied this method
to calculate fusion and capture cross sections above and below the barrier. So far, we have studied
the systems $^{132}$Sn+$^{64}$Ni~\cite{UO07a}, $^{64}$Ni+$^{64}$Ni,
$^{16}$O+$^{208}$Pb~\cite{UO09b}, $^{132,124}$Sn+$^{96}$Zr,
and we have studied the entrance channel dynamics of hot and cold fusion reactions leading
to superheavy element $Z=112$~\cite{UO10a}. Most recently, we have investigated 
sub-barrier fusion and pre-equilibrium giant resonance excitation between
various tin + calcium isotopes~\cite{OU12} and calcium + calcium isotopes~\cite{KU12}. Last not least, 
we have studied sub-barrier fusion reactions between both stable and neutron-rich
isotopes of oxygen and carbon~\cite{UO12} that occur in the neutron
star crust. In all cases, we have found
good agreement between the measured fusion cross sections and the DC-TDHF results.
This is rather remarkable given the fact that the only input in DC-TDHF is the 
Skyrme effective N-N interaction, and there are no adjustable parameters.
 

\section{Unrestricted TDHF dynamics}

The TDHF equations of motion are obtained from the variational principle
\begin{equation}
\delta S = \delta \int_{t_1}^{t_2} dt <\Phi(t) | H - i \hbar \frac{\partial}{\partial t} | \Phi(t) > = 0 \ ,
\label{eq:var_princ}
\end{equation}
where $H$ denotes the quantum many-body Hamiltonian of the system consisting of kinetic
energy and Coulomb / nuclear two-body interactions. The main approximation in TDHF is
that the many-body wave function $\Phi(t)$  is assumed to be a single time-dependent
Slater determinant which consists of an anti-symmetrized product of single-particle
wave functions
\begin{equation}
\Phi(r_1,...,r_A;t) = (A!)^{-1/2} \ det |\phi_{\lambda} (r_i,t) | \ .
\label{eq:Slater}
\end{equation}
The variational principle confined to the subspace of Slater determinants yields the TDHF equations
for the single-particle wave functions
\begin{equation}
h(\{\phi_{\mu}\}) \ \phi_{\lambda} (r,t) = i \hbar \frac{\partial}{\partial t} \phi_{\lambda} (r,t)
            \ \ \ \ (\lambda = 1,...,A) \ ,
\label{eq:TDHF}
\end{equation}
where $h = \partial E / \partial \rho$ denotes the mean-field Hamiltonian.

In the present TDHF calculations we use the Skyrme SLy4 interaction~\cite{CB98} for the nucleons
including all of the time-odd terms in the mean-field Hamiltonian~\cite{UO06}.
The numerical calculations are carried out on a 3D Cartesian lattice. For
$^{40,48}$Ca+$^{132}$Sn the lattice spans $50$~fm along the collision axis and $30-42$~fm in
the other two directions, depending on the impact parameter. Derivative operators on
the lattice are represented by the Basis-Spline collocation method. One of the major
advantages of this method is that we
may use a relatively large grid spacing of $1.0$~fm and nevertheless achieve high numerical
accuracy. First we generate very accurate static HF wave functions for the two nuclei on the
3D grid. The static HF equations are solved with the damped gradient iteration method.
The initial separation of the two nuclei is $22$~fm for central collisions. In the second
step, we apply a boost operator to the single-particle wave functions. The time-propagation
is carried out using a Taylor series expansion (up to orders $10-12$) of the unitary mean-field propagator,
with a time step $\Delta t = 0.4$~fm/c. 

\begin{figure}[h]
\begin{minipage}{18pc}
\includegraphics[width=18pc]{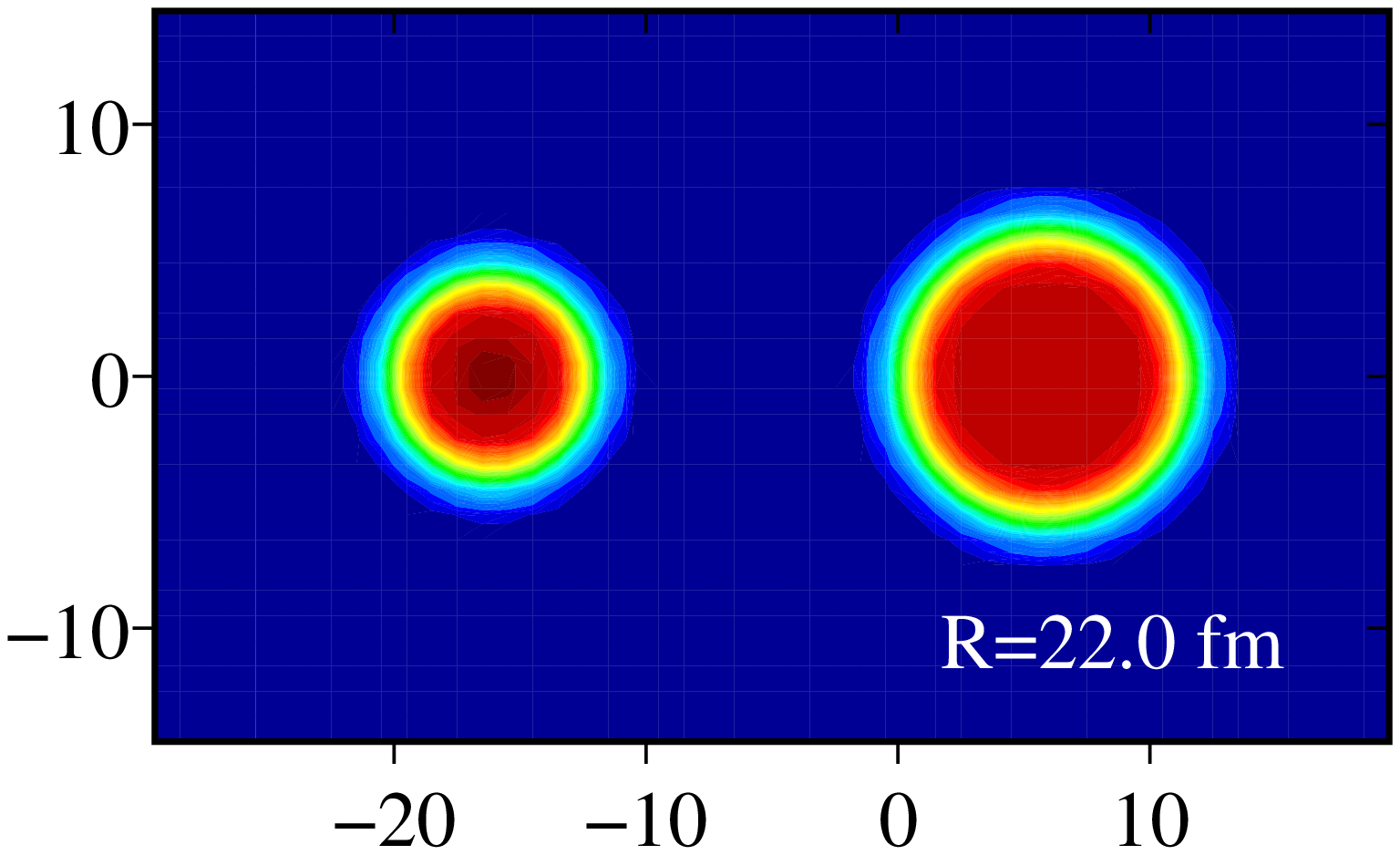}
\caption{\label{fig:dens22.0}Central collision of $^{48}$Ca+$^{132}$Sn
at $E_\mathrm{c.m.}=140$~MeV. Shown is a contour plot of the
TDHF mass density at internuclear distance $R=22.0$ fm.}
\end{minipage}\hspace{2pc}%
\begin{minipage}{18pc}
\includegraphics[width=18pc]{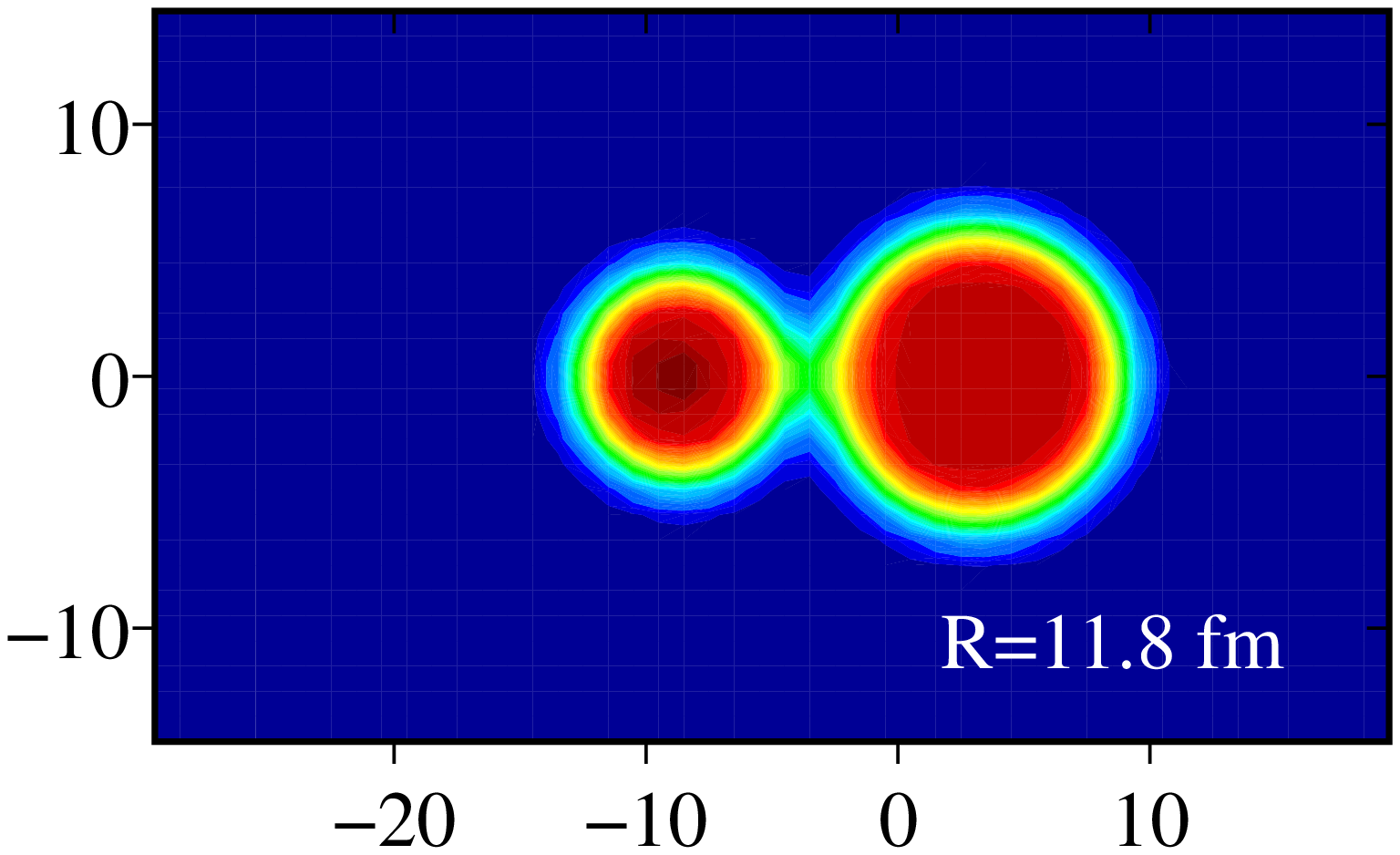}
\caption{\label{fig:dens11.8}Mass density at $R=11.8$ fm. A neck
between the two fragments is starting to develop.}
\end{minipage} 
\end{figure}

\begin{figure}[h]
\begin{minipage}{18pc}
\includegraphics[width=18pc]{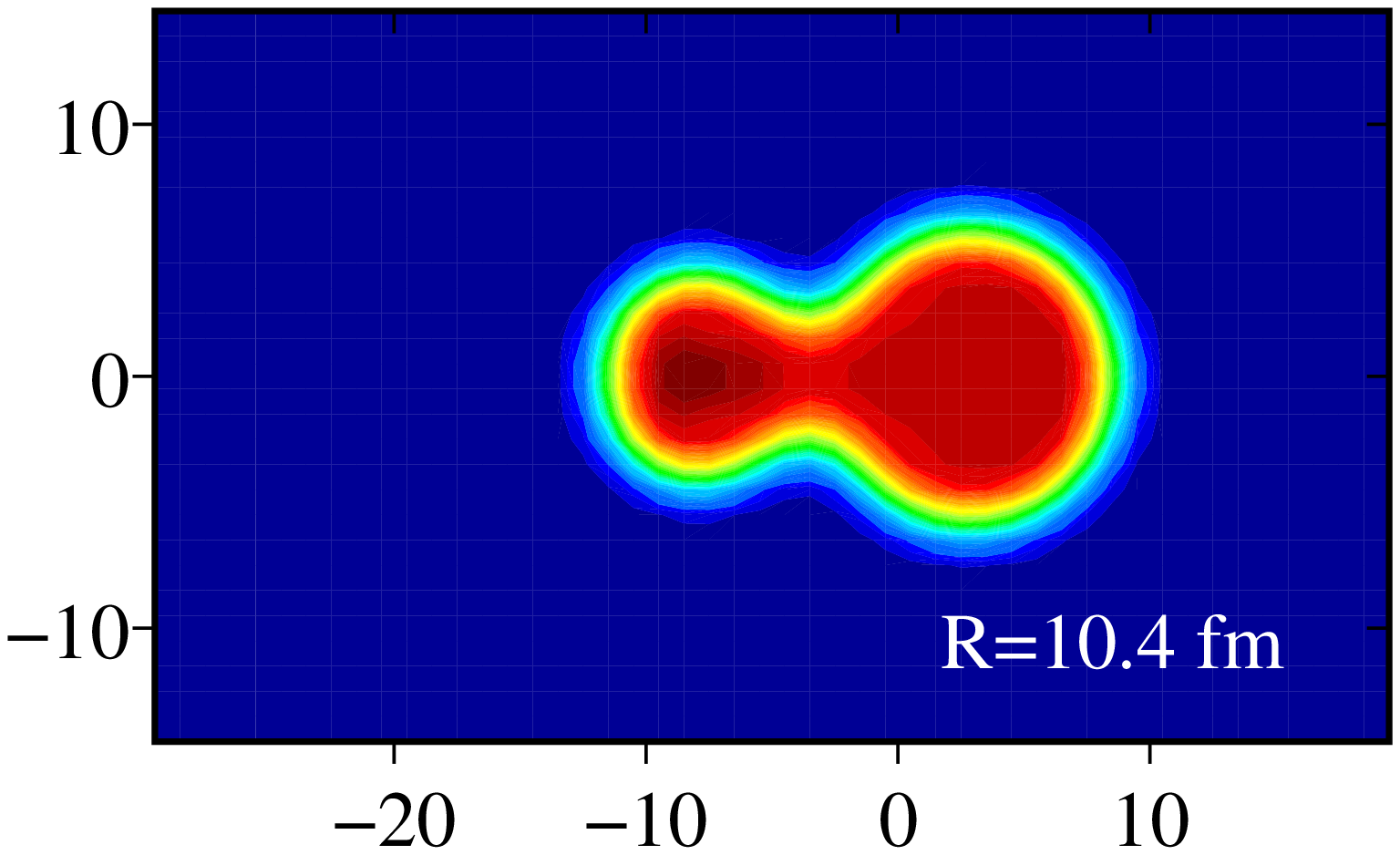}
\caption{\label{fig:dens10.4}Mass density at $R=10.4$ fm.
The neck broadens resulting in increased mass transfer.}
\end{minipage}\hspace{2pc}%
\begin{minipage}{18pc}
\includegraphics[width=18pc]{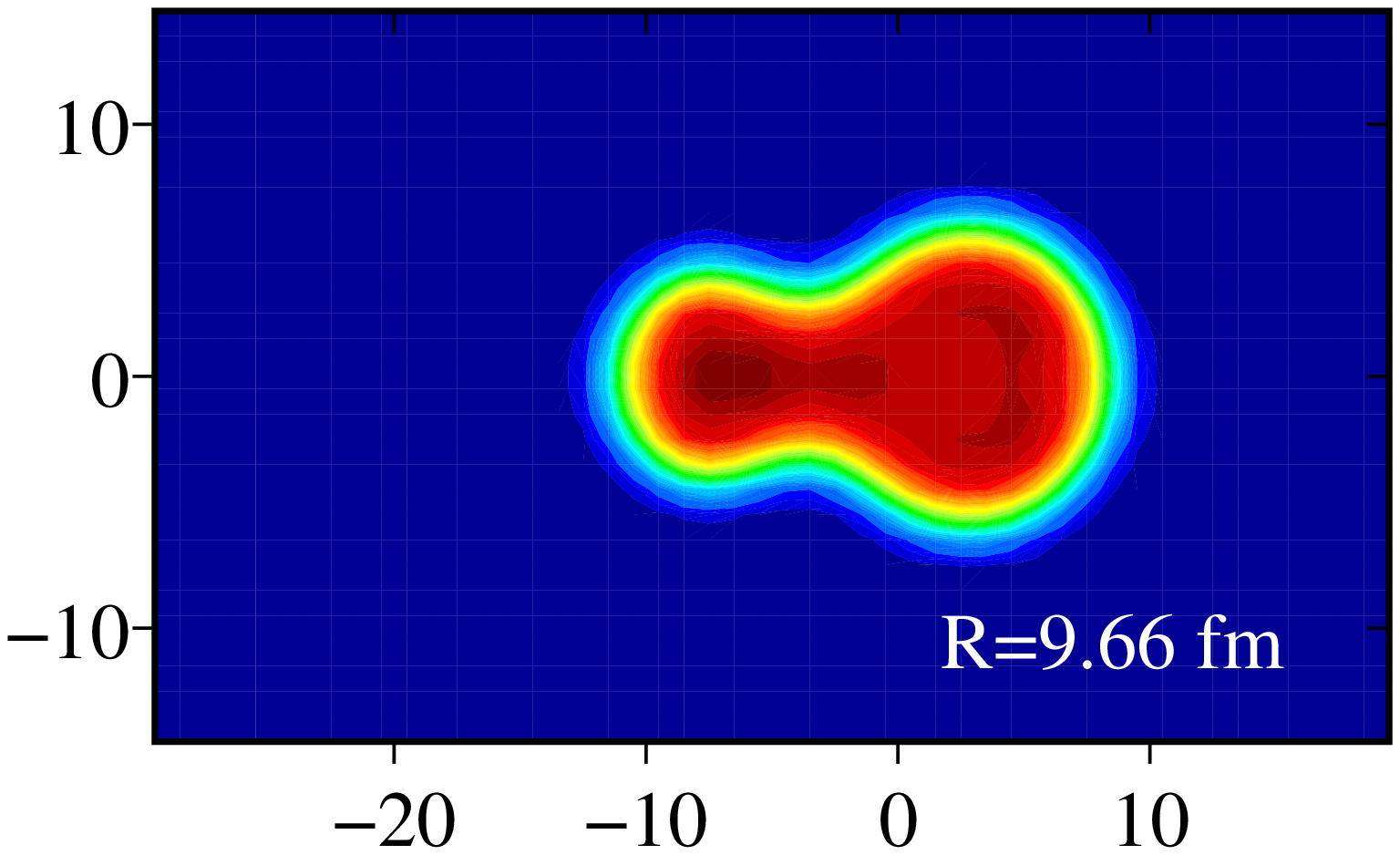}
\caption{\label{fig:dens9.66}Mass density at $R=9.66$ fm
corresponding to the capture point.}
\end{minipage} 
\end{figure}

In Figures~\ref{fig:dens22.0},\ref{fig:dens11.8},\ref{fig:dens10.4},\ref{fig:dens9.66} we show contour plots
of the TDHF mass density for a central collision of $^{48}$Ca+$^{132}$Sn
at $E_\mathrm{c.m.}=140$~MeV. The density distributions are shown as a function of
time, or equivalently, as a function of the internuclear distance $R$.

\section{DC-TDHF method: ion-ion potential and fusion cross section}

In the absence of a true quantum many-body theory of barrier tunneling, sub-barrier fusion
calculations are reduced to the calculation of a potential barrier between the interacting
nuclei and a subsequent calculation of tunneling through the barrier~\cite{UO07a}.
To date, theoretical studies of fusion cross sections are still dominated
by phenomenological methods such as the coupled-channels (CC) approach~\cite{EJ10}.
In this approach, one either uses empirical parameterizations of the ion-ion potential (e.g. Woods-Saxon)
or one calculates the potential with the double-folding method using
experimental or theoretical nuclear densities for projectile and target.
In the latter case, one relies on the ``frozen density'' or ``sudden'' approximation
in which the nuclear densities are unchanged during the computation of the ion-ion
potential as a function of the internuclear distance.
The frozen density approximation ignores dynamical effects such as neck formation
during the nuclear overlap. 
It has been demonstrated that for deep sub-barrier energies the inner part of the
potential barrier plays a very important role~\cite{IH07a}.
 
While phenomenological methods provide a useful starting point for the analysis
of fusion data, it is desirable to use a microscopic many-body theory.
We have developed a fully microscopic method to extract heavy-ion interaction potentials
$V(R)$ from the TDHF time-evolution of the dinuclear system.

In our DC-TDHF approach, the time-evolution takes place with no restrictions.
At certain times $t$ or, equivalently, at certain internuclear distances
$R(t)$ the instantaneous TDHF density 
\begin{equation}
\rho_{\mathrm{TDHF}}(r,t) = <\Phi(t) |\rho| \Phi(t) >
\label{eq:rho_TDHF}
\end{equation}
is used to perform a static Hartree-Fock energy minimization
\begin{equation}
\delta <\Phi_{\rho} \ | H - \int d^3r \ \lambda(r) \ \rho(r) \ | \Phi_{\rho} > = 0
\label{eq:var_dens}
\end{equation}
while constraining the proton and neutron densities to be equal to the instantaneous
TDHF densities
\begin{equation}
<\Phi_{\rho} |\rho| \Phi_{\rho} > = \rho_{\mathrm{TDHF}}(r,t) \ .
\label{eq:dens_constr}
\end{equation}
These equations determine the state vector $\Phi_{\rho}$. This means we
allow the single-particle wave functions to rearrange themselves in such a way
that the total energy is minimized, subject to the TDHF density constraint.
We have a self-organizing system which selects its path following
the dynamics given by the microscopic TDHF equations.

In a typical DC-TDHF run, we utilize a few
thousand time steps, and the density constraint is applied every $20$ time steps. 
We refer to the minimized energy as the ``density constrained energy'' $E_{\mathrm{DC}}(R)$
\begin{equation}
E_{\mathrm{DC}}(R) = <\Phi_{\rho} | H | \Phi_{\rho} > \ .
\label{eq:EDC}
\end{equation}
The ion-ion interaction potential $V(R)$ is essentially the same as $E_{\mathrm{DC}}(R)$,
except that it is renormalized by subtracting the constant binding energies
$E_{\mathrm{A_{1}}}$ and $E_{\mathrm{A_{2}}}$ of the two individual nuclei
\begin{equation}
V(R)=E_{\mathrm{DC}}(R)-E_{\mathrm{A_{1}}}-E_{\mathrm{A_{2}}}\ .
\label{eq:vr}
\end{equation}
The interaction potentials calculated with the DC-TDHF method incorporate
all of the dynamical entrance channel effects such as neck formation,
particle exchange, internal excitations, and deformation effects.
While the outer part of the potential barrier is largely determined by
the entrance channel properties, the inner part of the potential barrier
is strongly sensitive to dynamical
effects such as particle transfer and neck formation.

Using TDHF dynamics, it is also possible to compute the corresponding coordinate
dependent mass parameter $M(R)$ using energy conservation at zero impact parameter~\cite{UO09b}. 
As expected, at large distance $R$ the mass $M(R)$ is equal to the
reduced mass $\mu$ of the system. At smaller distances, when the nuclei overlap, the
mass parameter increases in all cases.
Instead of solving the Schr\"odinger equation with coordinate dependent
mass parameter $M(R)$ it is numerically 
advantageous to use the constant reduced mass $\mu$ and to transfer the
coordinate-dependence of the mass to a scaled
potential using a scale transformation (for details, see~\cite{UO09b}). For simplicity of
notation, we denote this transformed potential in the following by $V(R)$.
In general, we observe that the coordinate-dependent mass changes only 
the interior region of the potential barriers, and this change is most pronounced
at low $E_\mathrm{c.m.}$ energies.

The Schr\"odinger equation for the coordinate $R$ involving constant reduced mass $\mu$ and transformed
potential $V(R)$ has the familiar form
\begin{equation}
\left [ \frac{-\hbar^2}{2\mu} \frac{d^2}{dR^2} + \frac{\hbar^2 \ell (\ell+1)}{2 \mu R^2} + V(R)
 - E_\mathrm{c.m.} \right] \psi_{\ell}(R) = 0 \;.
\label{eq:Schroed1}
\end{equation}
By numerical integration of Eq.~(\ref{eq:Schroed1}) using the well-established
{\it Incoming Wave Boundary Condition} (IWBC) method we obtain
the barrier penetrabilities $T_{\ell}$ which determine the total fusion cross
section 
\begin{equation}
\sigma_{\mathrm{fus}}(E_{\mathrm{c.m.}}) = \frac{\pi \hbar^2}{2 \mu E_{\mathrm{c.m.}}}
                     \sum_{\ell=0}^{\infty} (2\ell+1) T_{\ell}(E_{\mathrm{c.m.}}) \ .
\label{eq:sigma_fus}
\end{equation}


\section{Numerical results}

\begin{figure}[h]
\begin{minipage}{18pc}
\includegraphics[width=18pc]{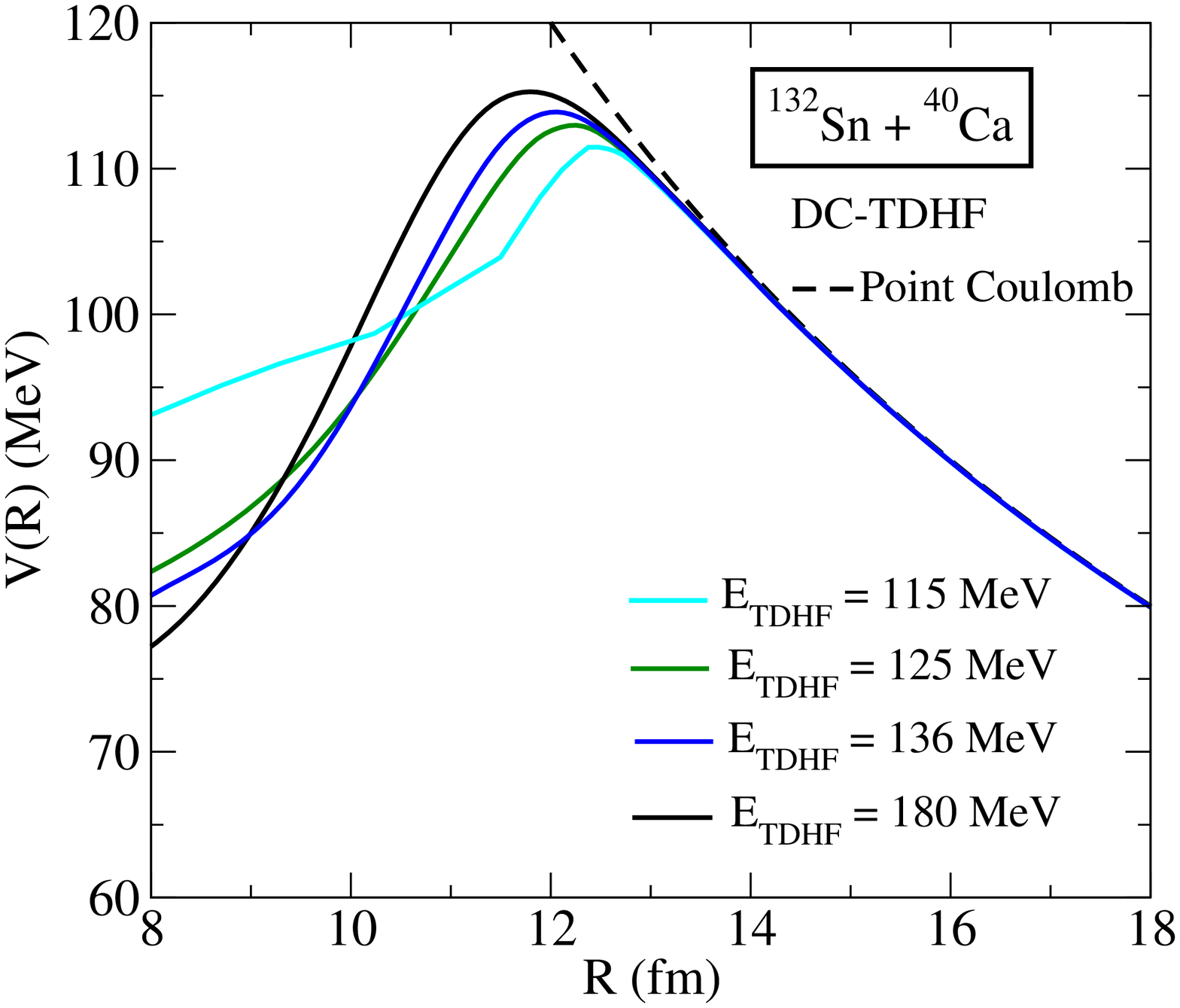}
\caption{\label{fig:pot_132Sn+40Ca_v3}Transformed heavy-ion interaction potentials
$V(R)$ corresponding to the reduced mass $\mu$. The potentials show a dramatic
dependence on energy.}
\end{minipage}\hspace{2pc}%
\begin{minipage}{18pc}
\includegraphics[width=18pc]{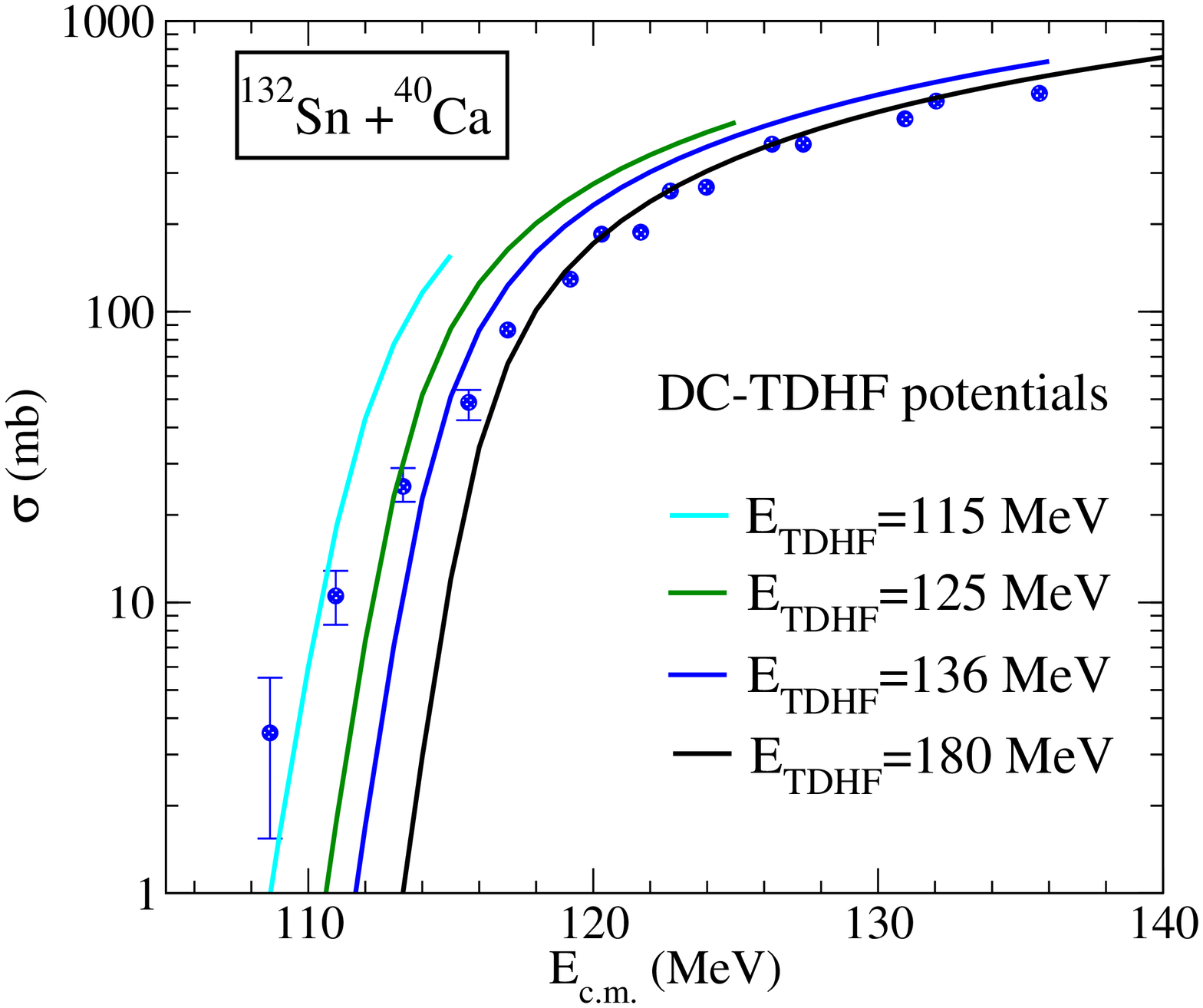}
\caption{\label{fig:sigma_fus_Sn132+Ca40}Total fusion cross sections obtained with the
DC-TDHF method for $^{132}$Sn+$^{40}$Ca. The cross 
sections are calculated from Eq.~\protect(\ref{eq:sigma_fus})
using the transformed potential $V(R)$ in Fig.~\ref{fig:pot_132Sn+40Ca_v3}
at four different energies. The experimental data are taken from Ref.~\cite{KR12}.}
\end{minipage} 
\end{figure}

In Fig.~\ref{fig:pot_132Sn+40Ca_v3} we show heavy-ion interaction potentials
for $^{132}$Sn+$^{40}$Ca calculated with the DC-TDHF method. The potentials
have been transformed so that they correspond to a constant reduced mass $\mu$.
In general, the potentials are energy-dependent and they are
shown here at four different TDHF energies. Our results demonstrate that in these heavy
systems the potential barrier height increases dramatically with increasing
energy $E_\mathrm{TDHF}$, and the barrier peak moves inward towards
smaller $R$-values. Note that the potential calculated at high energy
($E_\mathrm{TDHF}=180$~MeV) has a
barrier $E_B=115.3$ MeV located at $R=11.8$ fm,
whereas the potential calculated at low energy ($E_\mathrm{TDHF}=115$~MeV) has a
barrier of only $E_B=111.5$ MeV located at $R=12.4$ fm.

In Fig.~\ref{fig:sigma_fus_Sn132+Ca40} we show total fusion cross sections obtained with the
DC-TDHF method for $^{132}$Sn+$^{40}$Ca. Also shown are experimental data obtained at
HRIBF~\cite{KR12}. The main point of this display is to demonstrate
that the energy-dependence of the heavy-ion potential is crucial for an
understanding of the strong fusion enhancement at subbarrier energies.
At very high energy ($E_\mathrm{TDHF}=180$~MeV) the potential approaches
the limit of the frozen density approximation: the collision is so fast
that the nuclei have no time to rearrange their densities. 
We observe that the measured fusion cross sections at energies $E_\mathrm{c.m.}>118$~MeV
are well-described by this high-energy potential.
At low energies, however, the situation is quite different: Due to the slow motion of the nuclei,
the nuclear densities have time to rearrange resulting in neck formation, surface
vibrations, particle transfer, and the interior region of the heavy-ion potential is strongly modified.
These effects (which are included in TDHF) lower the fusion barrier
as described above, thus strongly enhancing the subbarrier fusion cross section.
We observe that the data points measured at the two lowest energies $E_\mathrm{c.m.}=108.6$ and $111$~MeV
can be understood by the heavy-ion potential calculated at $E_\mathrm{TDHF}=115$~MeV.

\begin{figure}[h]
\begin{minipage}{18pc}
\includegraphics[width=18pc]{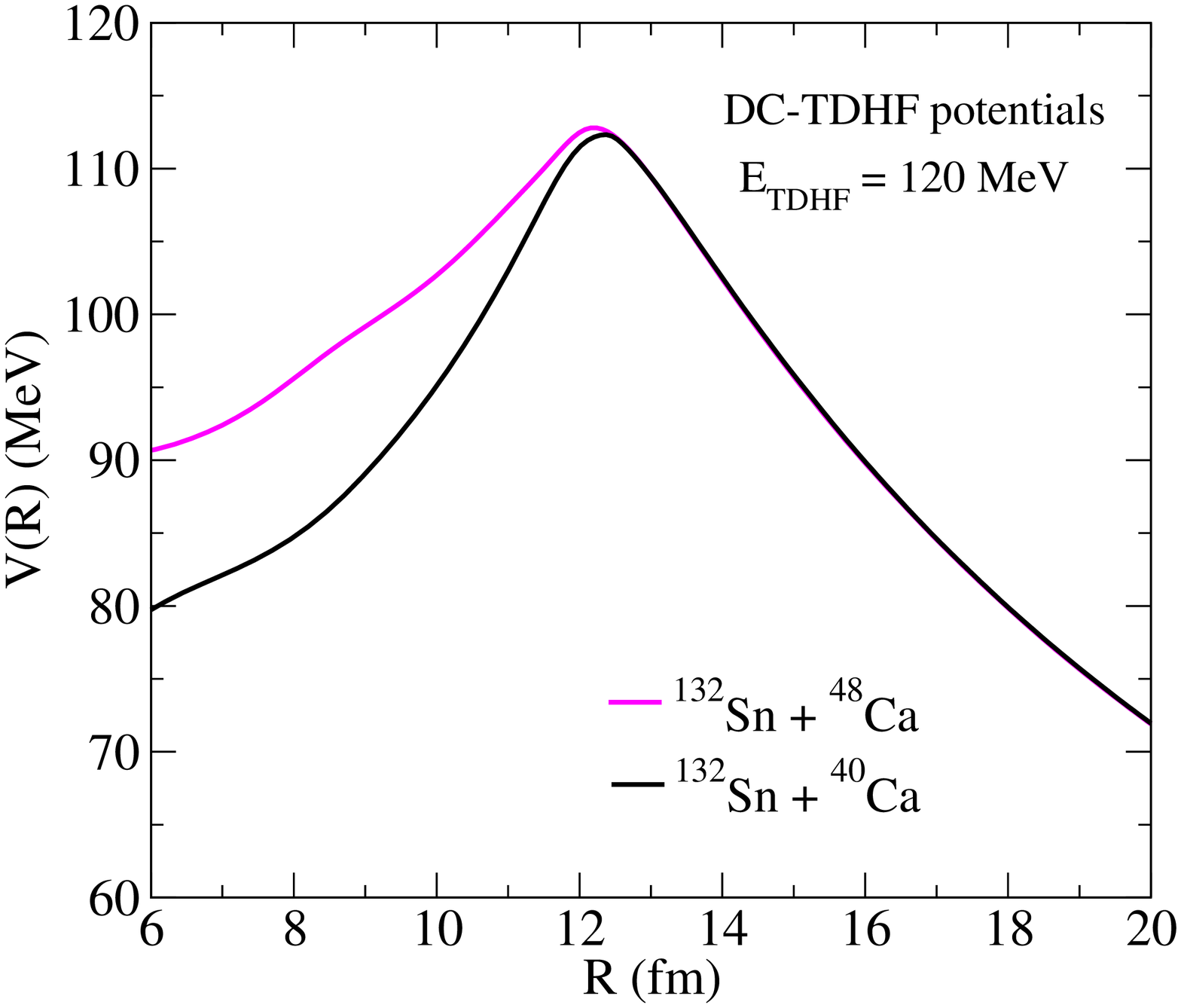}
\caption{\label{fig:pot2}Transformed heavy-ion potentials $V(R)$ at low energy
$E_\mathrm{TDHF}=120$~MeV for the systems $^{132}$Sn+$^{40,48}$Ca.}
\end{minipage}\hspace{2pc}%
\begin{minipage}{18pc}
\includegraphics[width=18pc]{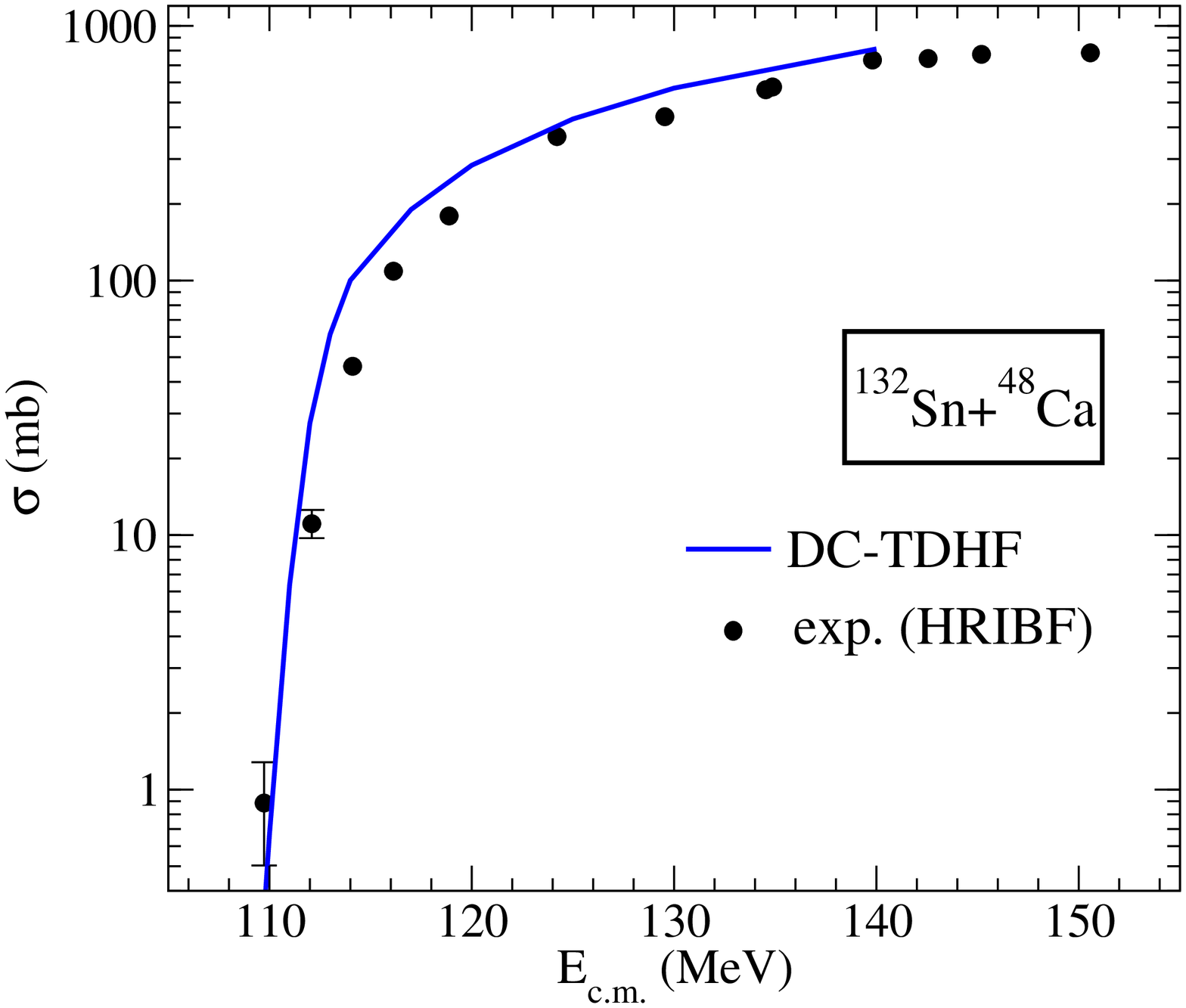}
\caption{\label{fig:sigma_fus_48Ca+132Sn}Total fusion cross sections obtained with the
DC-TDHF method for $^{132}$Sn+$^{48}$Ca.
The experimental data are taken from Ref.~\cite{KR12}.}
\end{minipage} 
\end{figure}
In Fig.~\ref{fig:sigma_fus_48Ca+132Sn} we show total fusion cross sections for $^{132}$Sn+$^{48}$Ca
which contains $8$ additional neutrons. In this case, we have interpolated the theoretical
cross sections obtained with the energy-dependent DC-TDHF potentials~\cite{OU12}. We can see that our
theoretical cross sections agree remarkably well with the experimental data. If one compares
the fusion cross sections for both systems at low energies, one finds the surprising result
that fusion of $^{132}$Sn with $^{40}$Ca yields a larger cross section than with $^{48}$Ca.
For example, at $E_\mathrm{c.m.}=110$ MeV we find an experimental cross section of $\approx 6$ mb for
$^{132}$Sn+$^{40}$Ca as compared to $0.8$ mb for the more neutron-rich system
$^{132}$Sn+$^{48}$Ca. This behavior can be
understood by examining the corresponding heavy-ion potentials which are shown in Fig.~\ref{fig:pot2}.
Both potentials have been calculated at the same center-of-mass energy $E_\mathrm{TDHF}=120$~MeV.
We observe that while the barrier heights and positions for both systems are approximately the same,
the \emph{width} of the DC-TDHF potential barrier for $^{132}$Sn+$^{40}$Ca is
substantially smaller than for $^{132}$Sn+$^{48}$Ca, resulting in enhanced sub-barrier fusion
at low energy.

\begin{figure}[h]
\begin{minipage}{18pc}
\includegraphics[width=18pc]{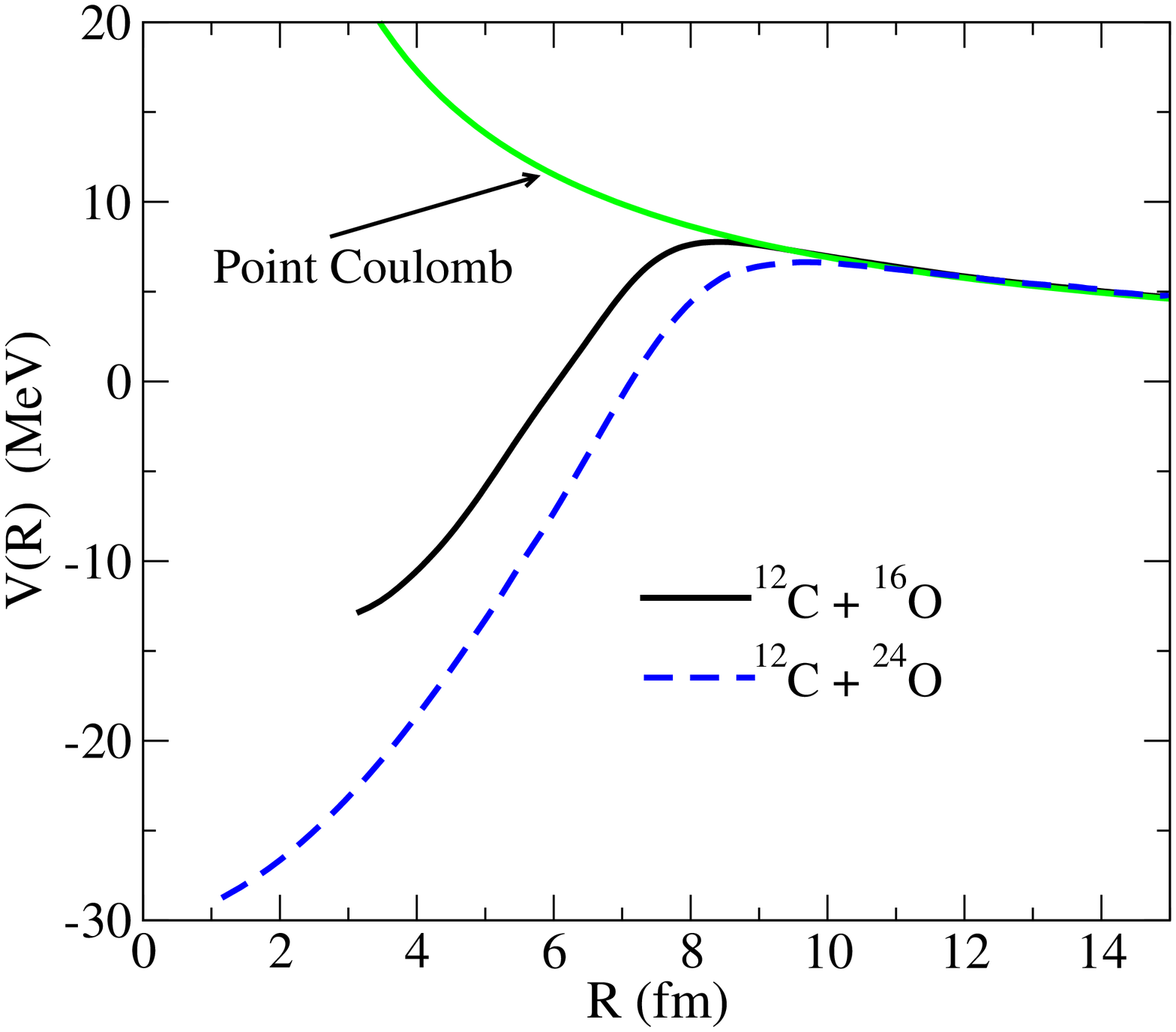}
\caption{\label{fig:pot_C+O}DC-TDHF heavy-ion potentials for the systems $^{12}$C+$^{16,24}$O.}
\end{minipage}\hspace{2pc}%
\begin{minipage}{18pc}
\includegraphics[width=18pc]{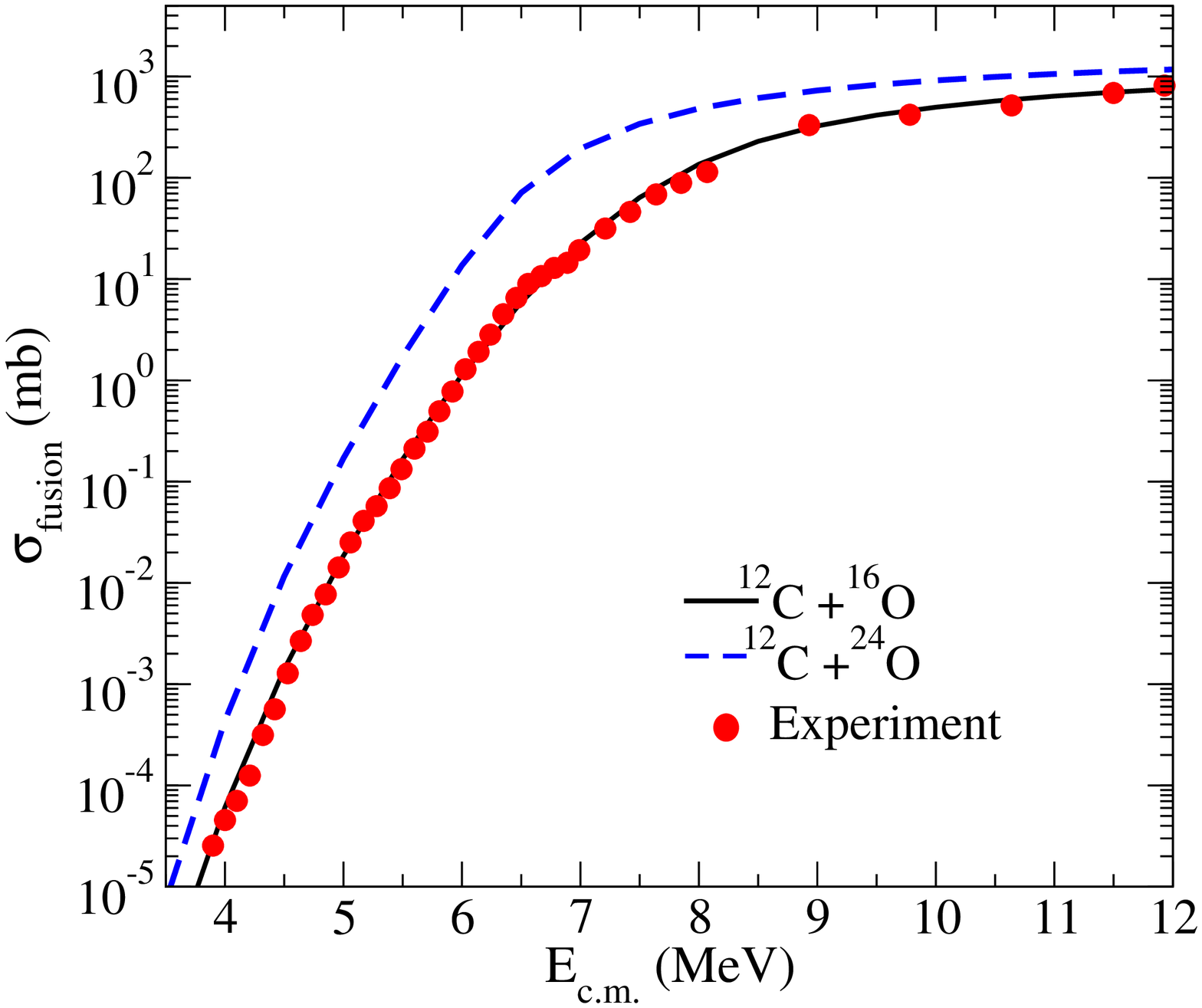}
\caption{\label{fig:sigma_fus_C+O}Total fusion cross sections versus center-of-mass
energy for fusion of carbon with oxygen isotopes. The experimental data are from Ref.~\cite{c12o16expdata_3}.}
\end{minipage} 
\end{figure}

Fusion of very neutron rich nuclei may be important to determine the composition and heating of the crust of accreting
neutron stars~\cite{UO12}. In Fig.~\ref{fig:pot_C+O} we show the DC-TDHF potential barriers for the C$+$O system.
The higher barrier corresponds to the $^{12}$C$+$ $^{16}$O system and has a peak
energy of $7.77$~MeV. The barrier for the $^{12}$C$+$ $^{24}$O system occurs at a
slightly larger $R$ value with a barrier peak of $6.64$~MeV.
Figure~\ref{fig:sigma_fus_C+O} shows the corresponding cross sections for the two reactions.
Also shown are the experimental data from Ref.~\cite{c12o16expdata_3}. The
DC-TDHF potential reproduces the experimental cross-sections quite well for the
$^{12}$C$+$ $^{16}$O system, and the cross section for the neutron rich $^{12}$C+$^{24}$O
is predicted to be larger than that for $^{12}$C +$^{16}$O.

We have compared our results to the phenomenological S\~{a}o Paulo barrier penetration model which
calculates an effective potential by folding over static densities for the projectile and target.
There are interesting differences between our calculations and the S\~{a}o Paulo model. For the asymmetric
systems $^{12}$C+$^{24}$O or $^{16}$O+$^{24}$O we predict an order of magnitude larger
cross section than for the S\~{a}o Paulo model. This is likely due to dynamical effects that change the nuclear
densities during the collision process caused by the rearrangement of the single-particle wave functions.
This enhancement of fusion cross sections of very neutron rich nuclei can be tested in the laboratory with
radioactive beams.

\section{Summary and conclusions}

There is mounting evidence that the TDHF equations give a good dynamical
description of the early stages of low-energy heavy-ion reactions.
We have developed powerful methods for extracting more information
from the TDHF dynamical evolution, e.g. the ion-ion interaction potential $V(R)$,
the coordinate-dependent mass parameter $M(R)$, and the dynamical excitation energy $E^*(R)$.
In general, $V(R)$ depends on the center-of-mass energy and, in the case of deformed nuclei,
on the Euler orientation angles.
The only input is the Skyrme N-N interaction, and there are no adjustable parameters.
We have calculated fusion cross sections below and above
the barrier for 18 different heavy-ion systems so far.
Overall, there is surprisingly good agreement with the experimental fusion data.
In the future, we believe that an effort is needed to incorporate deformation and scattering
information into the Skyrme N-N parametrization.


\ack
This work has been supported by the U.S. Department of Energy under Grant No.
DE-FG02-96ER40975 with Vanderbilt University. Some of the work presented here 
has been carried out in collaboration with J.A. Maruhn, P.-G. Reinhard, and C.J. Horowitz.
We also acknowledge fruitful discussions with our experimental colleagues J.F. Liang, J.J. Kolata,
W. Loveland and R.T. de Souza.



\section*{References}
\begin{thebibliography}{99}
\bibitem{KR12} Kolata J J, Roberts A, Howard A M, Shapira D, Liang J F, Gross C J, 
               Varner R L, Kohley Z, Villano A N, Amro H, Loveland W and Chavez E 2012
               {\it Phys. Rev.} C {\bf 85} 054603
\bibitem{desouza} Rudolph M J, Gosser Z Q, Brown K, Hudan S, de Souza R T,
                  Chbihi A, Jacquot B, Famiano M, Liang J F, Shapira D and Mercier D 2012
                  {\it Phys. Rev.} C {\bf 85} 024605 
\bibitem{Ne82}  Negele J W 1982 {\it Rev. Mod. Phys.} {\bf 54} 913 
\bibitem{Cus85a} Cusson R Y, Reinhard P G, Strayer M R, Maruhn J A and Greiner W 1985
                 {\it Z. Phys.} A {\bf 320} 475 
\bibitem{UO06}  Umar A S and Oberacker V E 2006 {\it Phys. Rev.} C {\bf 73} 054607 
\bibitem{KS10} Kedziora D J and Simenel C 2010 {\it Phys. Rev.} C {\bf 81} 044613 
\bibitem{Si11} Simenel C 2011 {\it Phys. Rev. Lett.} {\bf 106} 112502
\bibitem{GM08} Guo L, Maruhn J A, Reinhard P G and Hashimoto Y 2008 {\it Phys. Rev.} C {\bf 77}, 041301(R) 
\bibitem{DD-TDHF} Washiyama K and Lacroix D 2008 {\it Phys. Rev.} C {\bf 78} 024610 
\bibitem{CB98}  Chabanat E, Bonche P, Haensel P, Meyer J and Schaeffer R 1998
                {\it Nucl. Phys.} A {\bf 635} 231 
\bibitem{Klu09a} Kl\"upfel P, Reinhard P G,  B\"urvenich T J and Maruhn J A 2009
                {\it Phys. Rev.} C {\bf 79} 034310 
\bibitem{KL10}  Kortelainen M, Lesinski T, Mor\'{e} J, Nazarewicz W, Sarich J, Schunck N,
                Stoitsov M V and Wild S 2010 {\it Phys. Rev.} C {\bf 82} 024313 
\bibitem{UO06a} Umar A S and Oberacker V E 2006 {\it Phys. Rev.} C {\bf 74} 021601(R) 
\bibitem{UO07a} Umar A S and Oberacker V E 2007 {\it Phys. Rev.} C {\bf 76} 014614 
\bibitem{UO09b} Umar A S and Oberacker V E 2009 {\it Eur. Phys. J.} A {\bf 39} 243 
\bibitem{UO10a} Umar A S, Oberacker V E, Maruhn J A and Reinhard P G 2010
                {\it Phys. Rev.} C {\bf 81} 064607 
\bibitem{OU12} Oberacker V E, Umar A S, Maruhn J A and Reinhard P G 2012 {\it Phys. Rev.} C {\bf 85} 034609 
\bibitem{KU12} Keser R, Umar A S and Oberacker V E 2012 {\it Phys. Rev.} C {\bf 85} 044606 
\bibitem{UO12} Umar A S, Oberacker V E and Horowitz C J 2012 {\it Phys. Rev.} C {\bf 85} 055801
\bibitem{EJ10}  Esbensen H, Jiang C L and Stefanini A M 2010 {\it Phys. Rev.} C {\bf 82} 054621
\bibitem{IH07a} Ichikawa T, Hagino K and Iwamoto A 2007 {\it Phys. Rev.} C {\bf 75} 057603
\bibitem{HW07}  Hagino K and Watanabe Y 2007 {\it Phys. Rev.} C {\bf 76} 021601(R)
\bibitem{c12o16expdata_3} Jiang C L, Rehm K E , Back B B and Janssens R V F 2007
              {\it Phys. Rev.} C {\bf 75} 015803 
\end{thebibliography}
\end{document}